\documentclass[doublecol,figures]{epl2} 
\usepackage{amsmath}
\usepackage{amssymb}
\usepackage{amsfonts}
\makeatletter

\title{Spatiotemporal expressions reflecting topological classes of repressor networks}
\shorttitle{Spatiotemporal expressions reflecting topological classes of repressor networks} 

\author{Hiroki Ohta\inst{1,2,3} \and Mogens H. Jensen\inst{2}}
\shortauthor{H. Ohta \and M. H. Jensen}

\institute{                    
  \inst{1} Niels Bohr International Academy, 
University of Copenhagen, Blegdamsvej 17, Copenhagen 2100, Denmark\\
  \inst{2} Center for Models of Life, University of Copenhagen, Blegdamsvej 17, Copenhagen 2100, Denmark\\
  \inst{3} Department of Physics, Kyoto University, Kitashirakawa Oiwake-cho, Kyoto 606-8502, Japan\\
  }


\abstract{A family of repressor networks is proposed
  as a simple model of gene regulatory networks.
  We analytically show three topological classes of the repressor networks,
  each of which exhibits distinctly growing complexity of spatiotemporal expressions
  starting from nearly homogeneous states. Further, by focusing on locally interacting cases such as chain networks,
  including a generalized repressilator, or feedforward(back)-loop networks,
  spatiotemporal expressions in the long time regime and
  elusive relationships between such different networks are discussed in detail.}

\begin{document}
\date{\today}
\maketitle

\section{Introduction}
The accumulated data for cell dynamics,
obtained through rapidly developing experimental technology,
have inspired theoretical frameworks to
describe and analyze the complex dynamical behaviors in a simplified way \cite{Uri}.
One of the most striking achievements is the foundation of network motifs in gene regulatory networks,
i.e., the topological structures of a regulatory network,
which appear more frequently than those constructed by random processes without special bias \cite{Alon1,Lassig,Alon2}. 
Conversely, these theoretical developments have also stimulated experimental studies on cell dynamics
including gene regulatory networks. In particular, beyond merely observing existing regulatory networks,
it has become feasible to build up desired DNA-based regulatory networks \cite{Syn1,Syn2}.

One of the earliest examples for such synthetic DNA-based regulatory network
is the so-called repressilator, consisting of three repressors incorporated into a cell, where protein concentrations exhibit oscillations \cite{EL}.  
Further, various spatiotemporal patterns
have been realized in DNA-based regulatory networks in experiments \cite{OES,Rondelez1,Rondelez2}.
However, even for simplified versions of real cellular networks,
the theoretical understanding of how the topology of regulatory networks
influences their spatiotemporal expressions is limited to few cases.

Chemical reaction network theory
has already produced a long list of rigorous arguments
about stationary states of reaction networks \cite{Craciun,SAF,Elisenda2},
possibly including gene regulatory networks \cite{Ingalls}.
However, compared to the obtained results for stationary states,
this theoretical framework is less powerful
for understanding dynamical behaviors,
which is one of the main issues in gene regulatory networks.
Based on this literature, one way to develop our understanding of
gene regulatory networks is to propose a model of
simple gene regulatory networks,
which nevertheless exhibits rich dynamical behaviors,
and extract universal relationships between
the topological structures of such networks
and their spatiotemporal expressions.

On the strategy mentioned above, one starting point could be an extension of the repressilator,
for example, by taking into account a general cycle length \cite{Fraser,Grep1}
or its simplified limit \cite{Grep2}.
Indeed, some variants of the repressilator have been known to show rich dynamics
such as spatiotemporal oscillation, spatially incommensurate oscillation, and chaos \cite{Ojalvo,JKP}.
In this paper, as a more generalized extension,
we propose a family of repressor networks characterized by their topology
and provide general arguments for formulating three classes of
growing complexity of spatiotemporal expressions depending on the network topology.
Further, we analyze the dynamics in the late time regime for locally interacting cases such as a generalized repressilator or feedforward(back)-loop in more detail.
 
\section{Model}
Let us consider $L$ different chemical species
and denote the density of each chemical species $n\in\Lambda_L\equiv\{1,2,\cdots,L\}$
by a state variable $\rho_n\in\mathbb{R}_+$,
where $\mathbb{R}_+$ is the set of all non-negative real numbers. 
In the absence of interactions between chemical species,
each density evolves as a function of time by simply obeying $\partial_t \rho_n = c - \gamma \rho_n$,
where $c,\ \gamma\in\mathbb{R}_+$ is a production rate and a degradation rate, respectively.

\begin{figure}
\onefigure[width=4.5cm,clip]{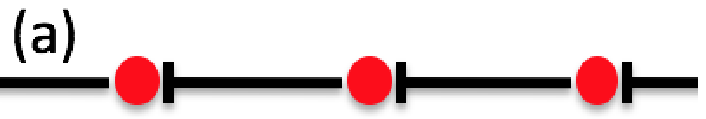}
\onefigure[width=4.5cm,clip]{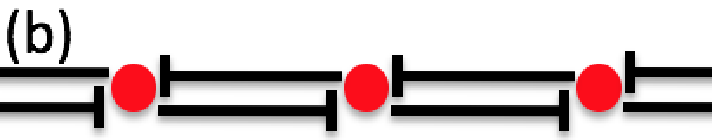}
\onefigure[width=4.5cm,clip]{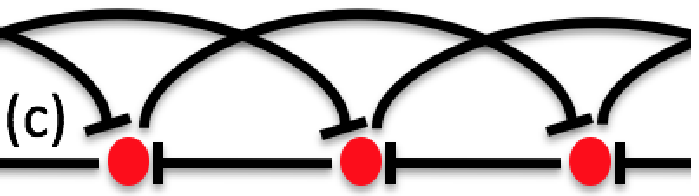}
\onefigure[width=4.5cm,clip]{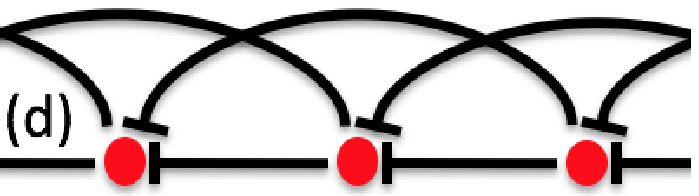}
\caption{(color online) Schematic pictures of repressor networks:
  Only three species among $L$ species are shown. For example,
  the species on the right represses the species at the center
  and the species at the center represses the species on the left.
(a) $\mathbb{C}=\mathbb{U}\equiv\{1\}$ (unidirectional chain). 
(b) $\mathbb{C}=\mathbb{B}\equiv\{1,-1\}$ (bidirectional chain).
(c) $\mathbb{C}=\mathbb{F}^-\equiv\{1,-2\}$ (feedback-loop network). 
(d) $\mathbb{C}=\mathbb{F}^+\equiv\{1,2\}$ (feedforward-loop network).}
\label{art}
\end{figure}

In order to express interactions between the different species on a network, 
we assign a set $\mathbb{C}\subseteq\mathbb{S}^L$, where $\mathbb{S}^L$ is
a set $\Lambda_{\lceil L/2\rceil-1}\cup \bar{\Lambda}_{\lceil L/2\rceil-1}\cup \{\lfloor L/2\rfloor\}$
with $\bar{\Lambda}_L \equiv\{-i\}_{i\in\Lambda_L}$. Some examples are shown in Fig. \ref{art}.
Here, we assume the standard form of a repressor interaction
characterized by Hill coefficient $h$ \cite{EL} and,
by using parameters of interaction strength $J,\ K_d\in\mathbb{R}_+$
with signed distance $d\in {\mathbb C}$, we consider the following model of a repressor network:
\begin{eqnarray}
\partial_t \rho_n = c - \gamma \rho_n + JF_{\rm int}(\{\rho_{n_d}\}_{d\in \mathbb C}),\label{model}\\
F_{\rm int}(\{\rho_{n_d}\}_{d\in\mathbb C})= \mathcal{N}^{-1}\sum_{d\in \mathbb C}\frac{K_d}{1+\rho_{n_{d}}^h}, 
\end{eqnarray} where $\mathcal{N}\equiv \sum_{d \in \mathbb C}K_d$
is the normalization for the interaction term $F_{\rm int}$ and $n_d \equiv n + d \ {\rm mod}\ L$;
namely, we impose periodic boundary conditions for simplicity.
Thus, this repressor network has translational invariance in terms of $n$.
Hereafter, we assume that $h>0$ and $K_d>0$ for any $d\in \mathbb C$,
and the network $(L,\mathbb{C})$ is irreducible, meaning that,
there is $d_p\in \{L\}\cup\mathbb{C}$ indivisible by $\min_{d\in\mathbb{C}\setminus \{1\}} d$.
For example, the chain $(L,\mathbb{C})=(8,\{2,4\})$ is reducible; this case corresponds
to the case of twice of the irreducible network $(4,\{1,2\})$.

This model has similarities to previously proposed models under some conditions.
Firstly, the network $(L,\mathbb {C})=(3,\{1\})$ corresponds
to the fast translation limit of repressilator \cite{EL}.
Secondly, suppose that a network with length $L=L_xL_y$
is divided into $L_x$-chains of length $L_y$
where the count of number begins with the bottom site of the left-most chain $1\in\Lambda_{L_x}$
toward the top site after $L_y-1$ steps and the count continues
from the bottom site of the next chain $2\in\Lambda_{L_x}$
towards the top site, and this process is repeated until $L$ is reached.
In this case, the network $(L,\mathbb C)=(L,\{-L_y,1,L_y-1\})$
corresponds to the repressor lattice \cite{JKP}, which we call $L_y$-layered repressor lattice in this paper.
Precisely speaking, the previously studied boundary conditions are
free boundaries or the standard periodic boundary conditions
instead of the skewed periodic condition as constructed above.

\section{Instability of homogeneous fixed points}
Let us begin with the case of $J=0$, meaning that there are no interactions.
In this case, it is trivial that the fixed point with
$\rho_n=\rho_{\rm s}(0)\equiv c/\gamma$ for any $n$ is globally stable.
Even in the case of $J>0$, it is straightforward to obtain a homogeneous fixed point
$\rho_n=\rho_{\rm s}(J)$ for any $n\in\Lambda_L$ satisfying the following relations:
\begin{eqnarray}
 & \rho_{\rm s}= R(\rho_{\rm s}),\label{eqhs}\\
 & R(x)=\gamma^{-1}\left( c+\dfrac{J}{1+x^h} \right).
\end{eqnarray} One may show that Eq. (\ref{eqhs})
has only one real solution $\rho_{\rm s}(J)$,
which is a monotonically increasing function of $J$ with,
at the leading order of $J$,
$\rho_{\rm s}(J)\to\rho_{\rm s}(0)+ \frac{J\gamma^{-1}}{(1+\rho_{\rm s}(0)^h)}$ as $J\to 0$
and $\rho_{\rm s}(J)\to (J\gamma^{-1})^{1/(h+1)}$ as $J\to\infty$.

Here, a natural question to ask is about the linear stability of
such a unique homogeneous fixed point
for a given condition and how its linear stability depends on the topological structures of the network.
Indeed, after linearizing the right-hand side of Eq. (\ref{model})
with $\rho_n(t) = \rho_{\rm s} + \delta\rho_n(t)$ and $\rho_{\rm s}\gg \delta\rho_n(t)$,
one may easily obtain the analytical expressions of eigenvalues
and eigenvectors of the obtained Jacobian matrix
because it is a circulant matrix \cite{eicir}. Specifically,
using the fact that the eigenvectors of a circulant matrix are discrete
Fourier modes with different wave numbers,
in order to compute the eigenvalues, one may formally assume
\begin{eqnarray}
\delta\rho_n(t)=\sum_{k\in\Lambda_L}A_k\exp\left(\lambda_kt + \boldsymbol{i} (2\pi k/L) n \right), \label{linb}
\end{eqnarray} where we have a complex number $\lambda_k$ as an eigenvalue,
the imaginary unit $\boldsymbol{i}$, and a real number $A_k$. 
Then, we immediately obtain the following equation determining the eigenvalues $\lambda_k$:
\begin{eqnarray}
\lambda_k = - \gamma - V(\rho_{\rm s}(J))M^{\mathbb C}_k,\label{eigen}\\
V(x)\equiv h\frac{(\gamma x - c)x^{h-1}}{(1+x^h)},\\
M^{\mathbb C}_k\equiv \mathcal{N}^{-1}\sum_{d\in \mathbb C} K_d\exp(\boldsymbol{i}(2\pi k/L)d),\label{Mf}
\end{eqnarray} where $-1\le |M^{\mathbb C}_k|\le 1$ \cite{Boun}. 
Indeed, one may easily show that $V(\rho_{\rm s})$ is a monotonically increasing function of $\rho_{\rm s}$
with $V(\rho_{\rm s}(0))=0$ and $V(\infty)=h\gamma$ because
$V'\equiv dV(x)/dx=V(x)\frac{h(1+x^h)-x}{x(1+x^h)} + \frac{cx^{h-2}}{1+x^h}> 0$
for $\rho_{\rm s}(0)\le x<\infty$ and $V'(\infty)=0$.

Next, we show the topological dependence of eigenvalues (\ref{eigen}) in terms of
$\{K_d\}_{d\in\mathbb C}$, which are directly related to the properties of $M^{\mathbb C}_k$.
Assume that $|\mathbb C|\ge 1$,
then the following properties hold for any $\{K_d\}_{d\in\mathbb C}$, and $(L,\gamma,c)$:

(0.i) {\it Existence of instability:}
There are certain finite values of $J$ and $h$,
above which the homogeneous fixed points are unstable. 
The instability point $J=J_{\rm c}(h)$ is explicitly defined by 
the smallest value of $J$ such that $\max_{k\in\Lambda_L}{\rm Re}(\lambda_k)\ge 0$,
where ${\rm Re}(x)$ is the real part of a complex number $x$.
Specifically, in order to have the instability,
$h>1$ is a necessary condition and $h > -1/\min_{k\in\Lambda} {\rm Re}(M^{\mathbb C}_k)$
is a sufficient condition because there is, at least, $k=k_-\in\Lambda_L$
such that ${\rm Re}(M^{\mathbb C}_{k_-})<0$.
This is simply because $\sum_{k\in\Lambda_L}M_k^{\mathbb C}=0$
leading to $\sum_{k\in\Lambda_{L-1}}M_k^{\mathbb C}=-M_L^{\mathbb C}<0$.

(0.ii) {\it Invariance of instability:}
Let us consider a flipping operator $f_d\in \{0,1\}$
in order to map $\{K_d\}_{d\in \mathbb C}$ into $\{K_d'\}_{d \in\mathbb C'}$,
where $K_{d}'= K_d(1-f_d)+K_{-d}f_{-d}$ for any $d\in\mathbb C$
and $\mathbb C'=\cup_{d\in{\mathbb C}}\{d-2f_dd\}$.
The instability point $J_c$ for the model
with any $\{K_d'\}_{d\in \mathbb C'}$ obtained by any possible flipping operators
is the same as that with a given $\{K_d\}_{d\in \mathbb C}$ as long as $J_{\rm c}$ exists
because of ${\rm Re}(\exp(\boldsymbol{i}x))={\rm Re}(\exp(-\boldsymbol{i}x))$.

\begin{figure}
  \onefigure[width=4.5cm,clip]{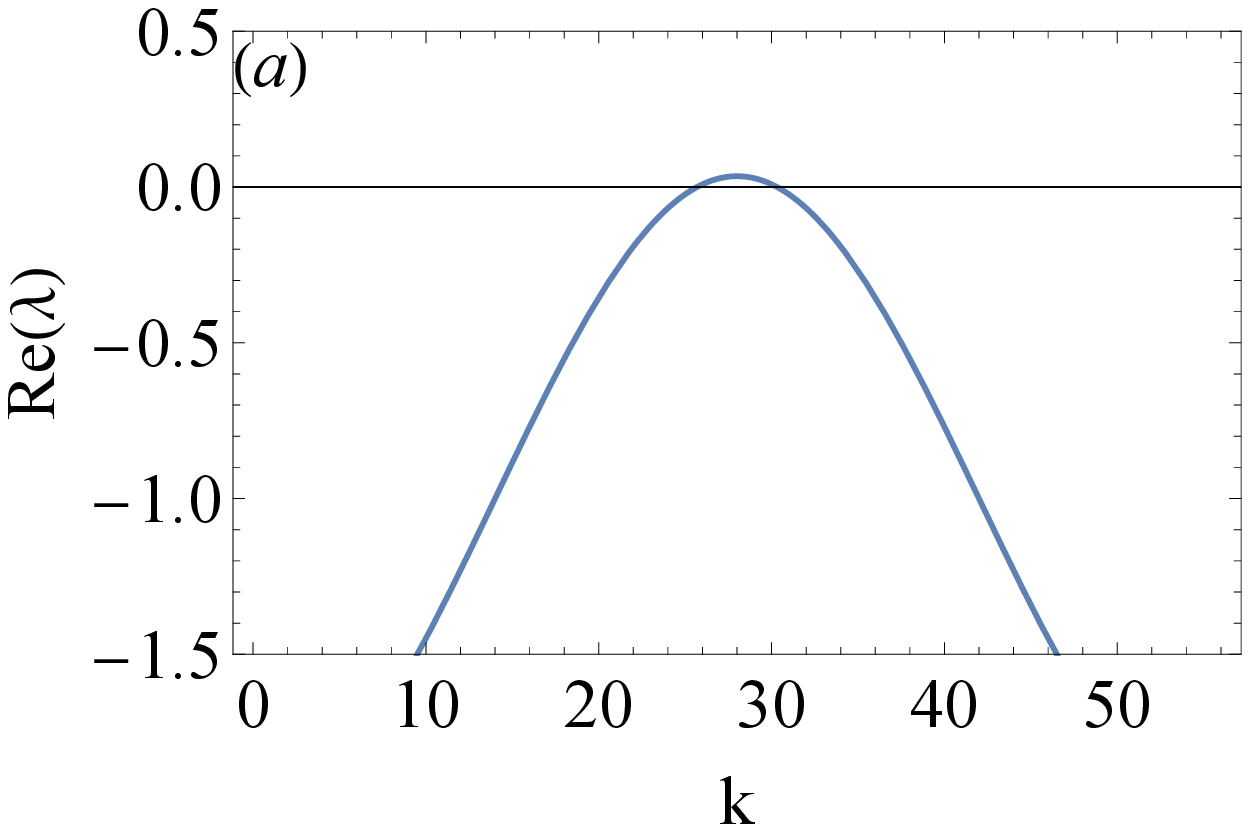}\onefigure[width=4.5cm,clip]{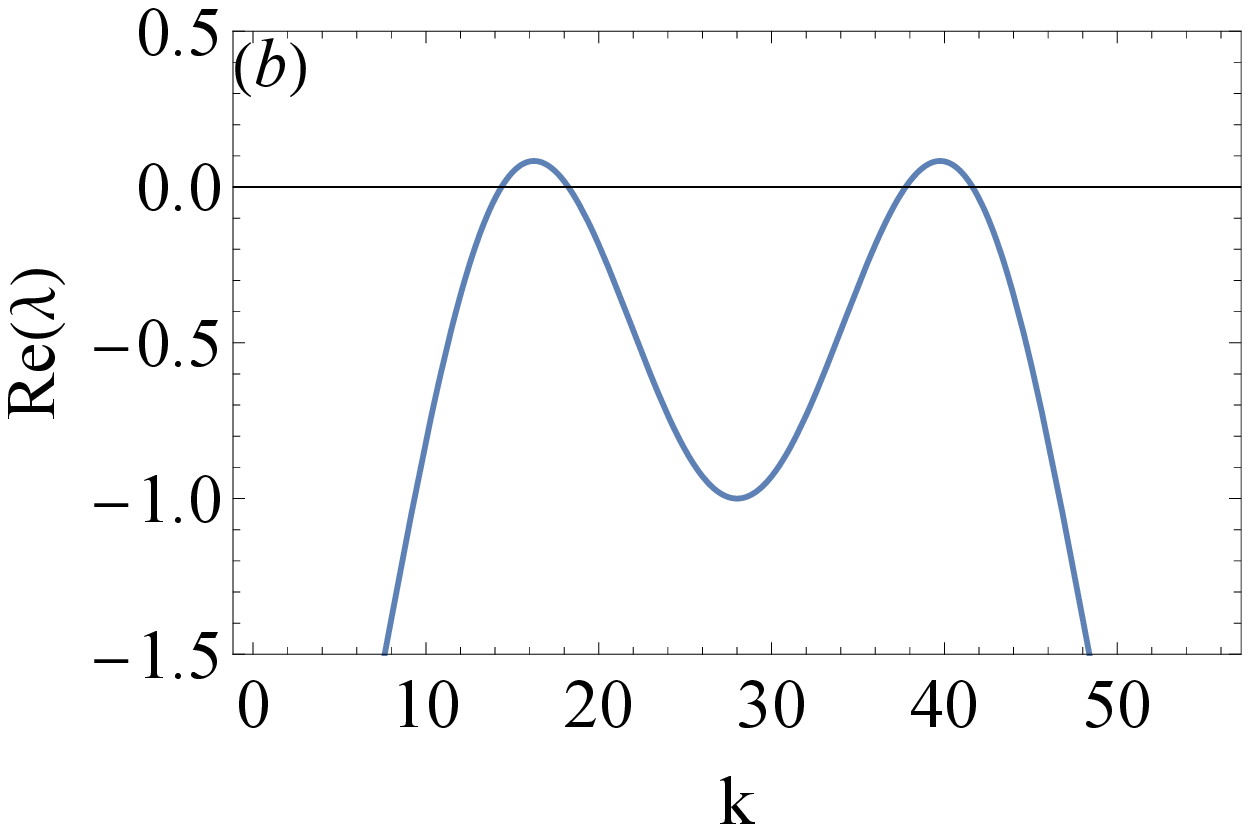}
  \caption{(color online) The real parts of eigenvalues $\lambda_k$, ${\rm Re}(\lambda)$,
    as a function of $k$ near instability point $J=J_{\rm c}$: 
  (a) $\mathbb{C}=\mathbb{U}$ (unidirectional chain)
  or $\mathbb{B}$ (bidirectional chain) with $J=1$. 
  (b) $\mathbb{C}=\mathbb{F}^-$ (feedback-loop network)
  or $\mathbb{F}^+$ (feedforward-loop network) with $J=2$. 
$L=56$, $h=4$, $\gamma=1$, and $c=0.1$.}
\label{eigenfig}
\end{figure}
Next, let us consider how the instability grows in the initial time regime
by focusing on the dominant eigenvalue,
which corresponds to the eigenvalue whose real part takes the maximum value
compared to the other eigenvalues.
Note that the dominant eigenvalues do not depend on $(J,h)$.
Hereafter, for convenience,
${\mathbb C}_{\rm o}$ denotes a set of all odd numbers in $\mathbb C$ and
${\mathbb C}_{\rm e}$ denotes a set of all even numbers in $\mathbb C$;
${\mathbb C}={\mathbb C}_{\rm o}\cup{\mathbb C}_{\rm e}$.
Indeed, one may derive the following arguments about the dominant eigenvalues
for certain classes of $\{K_d\}_{d\in \mathbb C}$, and $L$, which hold for any $(\gamma,c)$.

(1.i) {\it Monotonic mode:}
Suppose $K_{d}=K_{-d}$ holds for any $d\in \mathbb C\setminus\{L/2\}$.
Then, the Jacobian matrix in homogeneous fixed points is
symmetric and therefore all the eigenvalues are real numbers for any $(J,h,L)$.
Therefore, oscillatory behaviors are not expected to appear
in the initial time regime of the dynamics starting near a homogeneous fixed point.

For next two conditions (1.ii) and (1.iii), we assume that condition (1.i) is not satisfied.

(1.ii) {\it Zigzag mode:} 
Suppose $L$ is an even integer and $|{\mathbb C}_{\rm e}|=0$.
Then, the wave number of a dominant eigenvalue,
which we call $k_0$, is equal to $k_0=L/2$ for any $(J,h)$, 
and thus the imaginary part of this dominant eigenvalue is zero,
because of $\exp(\boldsymbol{i}d\pi) = -1$ for any odd $d$.
Note that in general, the other eigenvalues have non-zero imaginary parts.
This implies that a zigzag pattern with $\delta \rho_n(t) = -\delta\rho_{n+1}(t)$
is observed at the initial time regime of the instability.
Note that the instability point $J_{\rm c}$ satisfies $\gamma = V(\rho_{\rm s}(J_{\rm c}))$,
where $h>1$ is a sufficient condition to find a finite $J_{\rm c}$.

(1.iii) {\it Oscillatory mode:}
Suppose $L$ is either an odd integer or an even integer
with $\sum_{d\in{\mathbb C}_{\rm e}}K_{d} \ge \sum_{d'\in{\mathbb C}_{\rm o}}K_{d'}$.
Then, $k_0\neq L/2$ holds for any $(J,h)$ because of ${\rm Re}(M_{L/2}^{\mathbb C})\ge 0$,
meaning that there are two complex conjugate dominant eigenvalues with non-zero imaginary parts
unless $\sum_{d\in\mathbb C}K_d\sin((2\pi k_0/L) d)= 0$ is satisfied.
For example, if $|\mathbb C|=1$ and if $L$ and $d$ are coprime,
then the dominant eigenvalues always have a nonzero imaginary part at the instability point,
meaning that oscillatory behaviors appear in the initial time regime
near unstable homogeneous fixed points;
the case with $d=1$ corresponds to the generalized repressilator \cite{Grep1}.

Hereafter, in order to move on the dynamics in the late time regime,
we focus on locally interacting cases of networks
combined with numerical experiments, where the initial conditions obey
a Gaussian distribution with mean $\rho_s(0)=0.1$, variance $10^{-4}$, $L=56$,
and $h=4$ unless otherwise specified. Our numerical experiments show
that the spatiotemporal expressions do not qualitatively depend on $(L,h)$
as long as $L$ and $h$ are sufficient large,
except for special cases which will be pointed out later.

\section{Networks of chains}
\begin{figure}
\onefigure[width=6cm,clip]{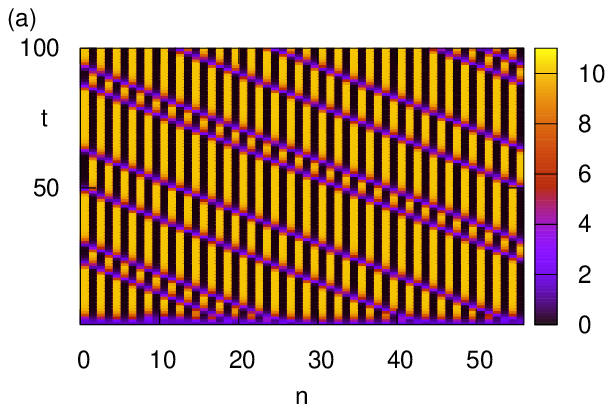}
\onefigure[width=6cm,clip]{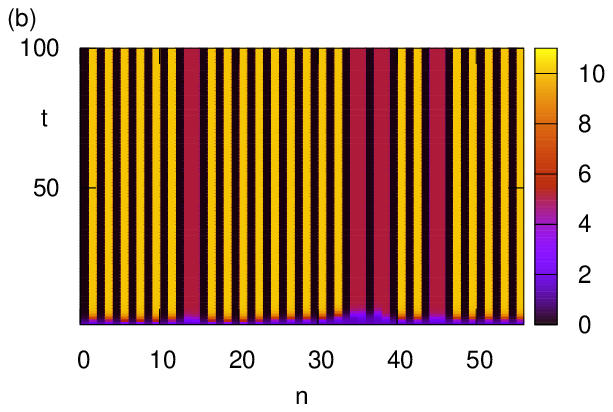}
\caption{(color online) Density map of $\rho_n$ as a function of $(n,t)$: 
  (a) Traveling point-defects in a crystal for ${\mathbb C}=\mathbb{U}$ (unidirectional chain). 
(b)  Point-defects in a crystal for ${\mathbb C}=\mathbb{B}$ (bidirectional chain).
  $L=56$, $h=4$, $\gamma=1$, $c=0.1$, and $J=10$.}
\label{E0fig}
\end{figure}
Let us consider a unidirectional chain $\mathbb{C}=\mathbb{U}\equiv\{1\}$
and a bidirectional chain $\mathbb{C}=\mathbb{B}\equiv\{1,-1\}$ with $K_{1} = K_{-1}$,
as shown in Fig. \ref{art}(a) and (b).
The case of $\mathbb{U}$ corresponds to property (1.ii) and
the generalized repressilator \cite{Grep1,Grep2}
where a {\it zigzag mode} may appear from the initial conditions
close to the homogeneous stationary solution.
On the other hand, the case with $\mathbb{B}$ corresponds to property (1.i)
where a {\it monotonic mode} may appear from such initial conditions.
Note that in both cases, the dominant eigenvalue behaves as shown in Fig. \ref{eigenfig}(a).
In order to elaborate on the dynamics in the long time regime,
we take into account the possibility of heterogeneous fixed points
of $c-\rho_n= F_ {\rm int}(\{\rho_{n_d}\}_{d\in \mathbb{C}})$ in both cases.
Indeed, for even $L$, one may find heterogeneous stationary solutions
$\rho_{n}=\rho_+$ and $\rho_{n_1}=\rho_-$ with $\rho_+\ge \rho_-$ for $n,n_1\in\Lambda_L$
such that $\rho_\pm = R(\rho_\mp)$ leading to
\begin{eqnarray}
\rho_\pm=R(R(\rho_\pm)).
\end{eqnarray} We call this fixed point a {\it zigzag solution} and
symbolically express it as $(+-+-)$.
Note that in the case of even $N$, there are two zigzag solutions $(+-+-)$ and $(-+-+)$,
where a one-step translational shift of $(+-+-)$ leads to $(-+-+)$.
Indeed, below the instability point $J=J_c$ of $\rho_s$,
$\rho_+=\rho_-=\rho_0$ always holds and above $J=J_c$,
$\rho_+>\rho_s>\rho_-$ may hold, as discussed in \cite{Grep1},
where $\rho_+(J)\to \frac{J\gamma^{-1}}{(1+\rho_{\rm s}(0)^h)}$
and $\rho_-(J)\to \rho_s(0)+(J\gamma^{-1})^{-(h-1)}(1+\rho_{\rm s}(0)^h)^h$
holds asymptotically, at the leading order of $J$, as $J\to\infty$;
Specifically, $\rho_+= 10.099..$, $\rho_-= 0.100961..$, and $\rho_0=1.556..$ for $J=10$.
Further, as also discussed in \cite{Grep1},
these zigzag solutions appearing after the instability of $\rho_{\rm s}$ are stable:
the eigenvalues are $-\gamma - \sqrt{V_0(\rho_+)(V_0(\rho_-))}\exp(\boldsymbol{i} (2\pi k/L))$
for $\mathbb{C}=\mathbb{U}$ and $-\gamma - \sqrt{V_0(\rho_+)(V_0(\rho_-))}\cos(2\pi k/L)$
for $\mathbb{C}=\mathbb{B}$ with $k\in\Lambda_{L}$
and $V_0(x)\equiv h\frac{Jx^{h-1}}{(1+x^h)^2}$ .

What our numerical experiments have shown for mainly $\mathbb{C}=\mathbb{U}$ are in the following.
For even $L$, in the initial time regime,
the dynamics from a homogeneous fixed point is expected to lead to the zigzag solution.
Contrary to this, as shown in Fig. \ref{E0fig}(a), an even number of traveling waves propagate
in a pattern possessing almost coinciding with a part of zigzag solutions 
in the long time regime, which can be pictured as traveling point-defects in a crystal.
Specifically, $\max_{n\in\Lambda_L}\rho_n\simeq 10.1$
and $\min_{n\in\Lambda_L}\rho_n\simeq 0.101$ for $J=10$
are consistent with a zigzag solution $\rho_+$ and $\rho_-$.
Further, each traveling defect shifts the local phase of the crystal,
where the speed of traveling waves $v$ depends on the parameter values \cite{proof}.
For odd $L$, a qualitatively similar behavior appears
but the number of traveling defects is odd, which is consistent with
the existence of the periodic attractor proven previously \cite{Smith,PSmith}.
Note that an even or odd number of traveling waves for even or odd $L$, respectively, are
commensurate with local patterns of the zigzag solution locating far from the traveling waves.
Thus, as $L\to\infty$, the perturbation from traveling waves to the local patterns of the zigzag solutions
would decrease as long as the number of traveling waves are odd or even integer
for odd or even $L$, respectively.
In this sense, the observed similar behaviors between for even and odd $L$
are not counterintuitive as long as $L$ is sufficiently large.
Note that the number and locations of traveling waves depend on the initial conditions.
Further, for the cases of $\mathbb{C}=\{1,d\}$ with $d\in\{3,-3,5,-5\}$,
there are also traveling waves moving to the direction of the sign of $-d$,
which implies that such behaviors observed in the case of $\mathbb{C}={\mathbb U}$
are rather universal for such networks with odd $d$.

For $\mathbb{C} =\mathbb{B}$, in addition to zigzag solutions,
it turns out that when $L$ is a multiple of three,
one may derive another set of heterogeneous stationary points
$\rho_{n}=\rho_{-'}$ and $\rho_{n_1}=\rho_{n_2}=\rho_{+'}$ with $\rho_{-'}<\rho_{+'}$ for $n,n_1,n_2\in\Lambda_L$
such that
\begin{eqnarray}
&\rho_{-'}=R(\rho_{+'}),\label{s1}\\
&\rho_{-'}=R(R(\rho_{-'})/2+\rho_{-'}/2).\label{s2}
\end{eqnarray} We call this fixed point a {\it spike solution} and express it symbolically as $(-'+'+'-')$,
where  $\rho_{+'}(J)\to \rho_+/2$
and $\rho_{-'}(J) \to \rho_{\rm s}(0) + (\rho_-(J) - \rho_{\rm s}(0))2^h$
holds asymptotically, at the leading order of $J$, as $J\to\infty$;
Specifically, $(\rho_{+'},\rho_{-'})=(5.10648..,0.11468..)$ for $J=10$.
Keeping these new heterogeneous fixed points in mind,
the numerical observations for $\mathbb{C}=\mathbb{B}$ are as follows.
First, a spatiotemporal pattern appears
from the instability of the homogeneous stationary solution as shown in Fig. \ref{E0fig}(b),
where $\max_{n\in\Lambda_L}\rho_n\simeq 10.1$ and $\min_{n\in\Lambda_L}\rho_n\simeq 0.101$
are very close to zigzag solutions with $\rho_+$ and $\rho_-$, respectively.
Second, there is a defect sequence $({\rm mppm})$ between the fragments of zigzag solutions,
where $\rho_{\rm p}\simeq 5.106$ and $\rho_{\rm m}\simeq 0.115$ when two defects are just adjacent
as $(\rm{ppmpp})$, which are very close to the spike solution $\rho_{+'}$ and $\rho_{-'}$, respectively;
otherwise $\rho_{\rm m}\simeq 0.107$.
Thus, by taking into account combinations of locating the fragments of spike solutions
$(-'+'+'-')$ in the fragements of zigzag solutions $(+-+-)$,
we conjecture that the number of locally stable heterogeneous solutions
for a network with length $L$ is, at least, $\sum_{k=1}^{L/4}{_{L/4}}C_{k}>2^{L/8}$.
The qualitatively similar behaviors,
which can be regarded as point-defects in a crystal, appear also for odd $L$.
Note that these heterogeneous solutions obtained in the long time limit depend on the initial conditions
and also there is another fixed point
$(\rho_{+'},\rho_{-'})=(1.372..,2.299..)$ satisfying Eqs. (\ref{s1}) and (\ref{s2}),
which, however, does not appear in the present condition presumably because it is unstable.

Furthermore, we have numerically investigated the dependence of spatiotemporal expressions
on $\alpha\equiv K_1/K_{-1}$ in the case of $\mathbb{C}=\mathbb{B}$.
Indeed, as long as $\alpha$ is sufficiently close to $1$,
the point-defects in a crystal are robust against changes in $\alpha$.
However, as $\alpha$ is decreased, at a certain value of $\alpha=\alpha_c$,
the defects becomes depinned, leading to traveling point-defects
in a crystal as observed in the case of $\mathbb{C}=\mathbb{U}$,
implying that there is a bifurcation between two qualitatively different behaviors.
It seems that these behaviors do not depend qualitatively on
whether $L$ is even or odd, as long as $L$ is sufficiently large.

\section{Networks of feedforward or feedback loops}
\begin{figure}
\onefigure[width=6cm,clip]{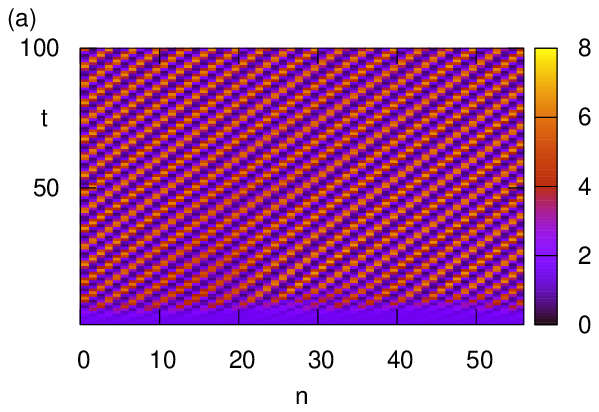}
\onefigure[width=6cm,clip]{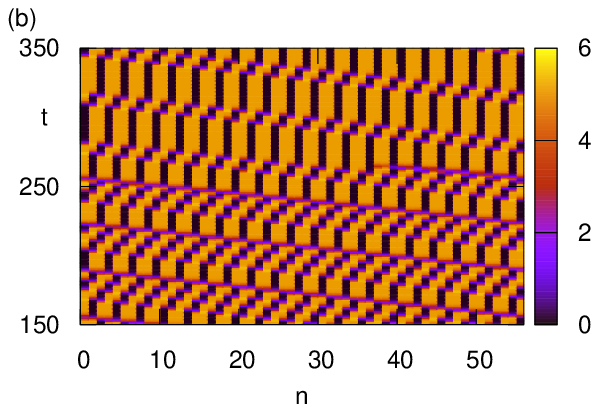}
\caption{(color online) Density map of $\rho_n$ as a function of $(n,t)$: 
(a) A traveling crystal for ${\mathbb C}=\mathbb{F^-}$ (feedback-loop network). 
  (b) Two kinds of traveling defects and its pair annihilation
  around $(n,t)=(40,270)$ in a crystal
  for ${\mathbb C}=\mathbb{F^+}$ (feedforward-loop network).
  $L=56$, $h=4$, $\gamma=1$, $c=0.1$, and $J=10$.}
\label{E1fig}
\end{figure}
We move onto the case of $\mathbb{C}=\mathbb{F}^\pm\equiv\{1,\pm 2\}$
with $K_1=K_{\pm 2}$ as shown in Fig. \ref{art}(c) and (d)
where $\mathbb{F}^+$ and $\mathbb{F}^-$ are regarded
as the network of minimal feedforward loops and feedback loops, respectively. 
This case corresponds to property (1.iii), where an {\it oscillatory mode} may appear
from the initial conditions close to the homogeneous stationary solutions,
as shown in Fig. \ref{eigenfig}(b). Furthermore, the feedback-loop-like case ${\mathbb F}^{-}$
corresponds to $L_y$-layered repressor lattice with $L_y=2$ \cite{JKP}.

In this case, at the instability point of $\rho_{\rm s}$,
the real part of the eigenvalue with $k=k_0$
(which is very close to $2L/7$) first becomes zero.
Note that the wavenumber of its complex conjugate is $k_*=L-k_0\simeq 5L/7$.
Thus, the spatiotemporal expressions in the initial time regime
can be described by $\rho_n(t)-\rho_{\rm s}$ proportional to $\cos(\omega_{\pm}t + n 2\pi k_0/L +\delta)$
with $\omega_\pm=2^{-1}V(\rho_{\rm s} (J))(\sin(2\pi k_0/L)+\sin(d_\pm2\pi k_0/L))$
where $\delta$ is a constant and $\mathbb{F}^\pm=\{1,d_\pm\}$.
In the numerical experiments for $\mathbb{C}=\mathbb{F}^-$ as shown in Fig. \ref{E1fig}(a),
even in the long time limit,
the similar patterns of traveling crystal expected from the above appear,
which do not depend on the initial conditions except for a spatiotemporal time shift
corresponding to the values of $\delta$ \cite{Note1}.
It should be pointed out that the behaviours similar to such a traveling crystal appear also
for $\mathbb{C}=\{1,d\}$ with $d\in\{4,6,-4,-6\}$.
It implies that such behaviors observed in the case of $\mathbb{C}={\mathbb F}^-$
are rather universal for those networks with even $d$.

On the other hand, the numerical observations for
the feedforward-loop-like case of ${\mathbb C}=\mathbb{F}^+$ are as follows.
Spatiotemporal patterns get more complicated in the late time regime
through transient dynamics as shown in Fig. \ref{E1fig}(b). 
In order to understand the late time regime, it turns out to be important to realize that spike solutions obtained
as $(-'+'+'-')$ symbolically for $\mathbb{C}=\mathbb{B}$
are stationary solutions also for $\mathbb{C}=\mathbb{F^+}$.
Note that when $L$ is a multiple of three, then the number of these spike solutions
is exactly three. Indeed, $\max_{n\in\Lambda_L}\rho_n\simeq 5.106$ and $\min_{n\in\Lambda_L}\rho_n\simeq 0.115$
are very close to the spike solution with $\rho_{+'}$ and $\rho_{-'}$, respectively.
Further, there are two kinds of traveling waves in the fragments of spike solutions, which
can be pictured as traveling defects in a crystal;
each traveling defect shifts the local phase of a crystal to the opposite directions.
Indeed, the speed of traveling waves causing the negative shift (negative wave) is
relatively faster than that of another causing the positive shift (positive wave),
leading to a pair annihilation when they collide.
Thus, in the long time limit, depending on the initial conditions,
either positive waves or negative waves remain,
whose number is $n_{\rm p}=2$ or $n_{\rm m}=1$, respectively for $L=56$.
The number of the traveling waves depends on the value of $L$
in such a way that $n_{\rm p}=1$ or $n_{\rm m}=2$ for $L=55$,
and $n_{\rm p}=3$ or $n_{\rm m}=0$ for $L=54$, where $n_{\rm m}=0$
means the existence of exact spike solutions with no negative waves
because $L=54$ is a multiple of three. 
Thus, if the number of traveling waves are commensurate with spike solutions as found above,
the traveling waves are stable in the long time scale.

Lastly, it turns out that 
in the case of the network $(L,\mathbb{C})=(56,\{1,6\})$,
there is a special heterogeneous stationary solution obtained approximately
by the repetition of sequence $(-'+'+'-'+-+-')$,
which exactly corresponds to a combination of the fragments of both spike and zigzag solutions
in the case of $\mathbb{C}=\mathbb{B}$.
The numerical experiments indicate
that whether a traveling crystal as observed for $\mathbb{C}=\mathbb{F}^-$ or such a special stationary solution
appears depend on the initial conditions.
This construction of special stationary solutions is available
as long as the number of repetition $L/(d+1)$ is an integer; particularly $L/(d+1)=8$ in this case.
Thus, by construction, these heterogeneous stationary solutions exist for any networks $(L,\mathbb{F}^{d})$
with $\mathbb{F}^{d}\equiv\{1,d\}$ such that $L/(d+1)$ with even $d\ge 6$ is an integer,
and are expected to be locally stable.

\section{Concluding remarks}
We have proposed a family of repressor networks
as an extension of the previously studied repressor networks \cite{Grep2,JKP}.
We have classified those networks into several topological classes
from the viewpoint of the instability of the homogeneous states
and clarified the mechanism of spatiotemporal expressions in the long time regime
for locally interacting cases.

Let us comment on the relationship between the present model
and the generalized repressilator previously proposed in the case of $\mathbb{U}$
as an extension of the original repressilator with finite translation speed \cite{Grep1}.
Indeed, by explicitly introducing another variable describing a finite translation speed
in the way of $\tau\partial_t\eta_n = \rho_n-\eta_n$ and replacing the interaction term by
$F_{\rm int}(\{\eta_{n_d}\}_{d\in\mathbb{C}})$, we have numerically found that
the spatiotemporal expressions observed in this paper do not change qualitatively,
at least, for networks of chains and feedforward(back)-loop, provided that $\tau$ is sufficiently small.
In particular, even in the presence of non-zero $\tau$ for $\mathbb{C}=\mathbb{U}$,
we have still observed traveling waves.

As a future study, it would be interesting to consider
the possibility of classifying the repressor networks
by taking into account the dynamics in the long time regime.
We have already found elusive relationships between the dynamics of different networks in the late time regime.
For example, spatiotemporal expression in $\mathbb{B}$ and $\mathbb{F}^{+}$
are closely related to each other, which might be classified into a subclass including other networks.
Further, it is a natural question how heterogeneity of chains and adding activators instead of repressors
will change the spatiotemporal behaviors as studied in \cite{Jensen2,Jensen3}.
Such studies based on the repressor networks potentially
shed light on the role of topological structures in gene regulatory networks 
and also how to experimentally design gene regulation networks
to control biological properties in a desired manner.


\acknowledgements
This work is supported by a Danish fund to Center for Models of life.
We thank C. Tian for comments about the reducibility of networks
and A. Dechant for a critical reading of this paper.
  H. O. thank E. Feliu and C. Wiuf for discussions
  on the related reaction network theory during his stay at their group as a visitor.


\begin{thebibliography}{99}
\bibitem{Uri}
 \Name{Alon U.}
  \Book{An Introduction to Systems Biology}
  \Publ{CRC Press}
  \Year{2006}.
\bibitem{Alon1}
  \Name{Milo R., Shen-Orr S., Itzkovitz S., Kashtan N., Chklovskii D., \and Alon U.}
  \REVIEW{Science}{298}{2002}{824}.
\bibitem{Lassig}
   \Name{Berg J. \and Lassig M.}
  \REVIEW{Proc. Natl. Acad. Sci.}{101}{2004}{14689}.
\bibitem{Alon2}
   \Name{Kashtan N., Itzkovitz S., Milo R., \and Alon U.}
  \REVIEW{Phys. Rev. E}{70}{2004}{031909}.
\bibitem{Syn1}
   \Name{Kim J. \and Winfree E.}
  \REVIEW{Mol. Syst. Biol.}{7}{2011}{465}.
\bibitem{Syn2} 
   \Name{Montagne K., Plasson R., Sakai Y., Fujii T., \and Rondelez Y.}
  \REVIEW{Mol. Syst. Biol.}{7}{2011}{466}.
\bibitem{EL}
   \Name{Elowitz M. B. \and Leibler S.}
  \REVIEW{Nature}{403}{2000}{335}.
\bibitem{OES}
   \Name{Garcia-Ojalvo J., Elowitz M. B., \and Strogatz S. H.}
  \REVIEW{Proc. Natl. Acad. Sci.}{101}{2004}{10955}.
\bibitem{Rondelez1}
   \Name{Genot A. J., Fujii T., \and Rondelez Y.}
  \REVIEW{Phys. Rev. Lett.}{109}{2012}{208102}.
\bibitem{Rondelez2}
   \Name{Zadorin A. S., Rondelez Y., Galas J. C., \and Torres A. E.}
  \REVIEW{Phys. Rev. Lett.}{114}{2015}{068301}.
\bibitem{Craciun}
  \Name{Craciun G., Tang Y., \and Feinberg M.}
  \REVIEW{Proc. Natl. Acad. Sci.}{103}{2006}{8697}.
\bibitem{SAF}
  \Name{Shinar G., Alon U., \and Feinberg M.}
  \REVIEW{SIAM J. Appl. Math.}{69}{2009}{977}.
\bibitem{Elisenda2}
   \Name{Feliu E. \and Wiuf C.} 
   \REVIEW{BMC Systems Biology}{9}{2015}{22}.
   \bibitem{Ingalls} 
  \Name{Ingalls B.}
  \Book{Mathematical Modelling in Systems Biology}
  \Publ{MIT press}
  \Year{2013}.
\bibitem{Fraser}
   \Name{Fraser A. \and Tiwari J.}
  \REVIEW{J. Theor. Biol.}{47}{1974}{397}.
\bibitem{Grep1} 
   \Name{Muller S., Hofbauer J., Endler L., Flamm C., Widder S., \and Schuster P.}
  \REVIEW{J. Math. Biol.}{53}{2006}{905}.
\bibitem{Grep2} 
   \Name{Pigolotti S., Krishna S., \and Jensen M. H.}
  \REVIEW{Proc. Natl. Acad. Sci.}{104}{2007}{6533}.
\bibitem{Ojalvo}
   \Name{Ullner E., Koseska A., Kurths J., Volkov E., Kantz H., \and Ojalvo J. G.}
  \REVIEW{Phys. Rev. E}{78}{2008}{031904}.
\bibitem{JKP}
   \Name{Jensen M. H., Krishna S., \and Pigolotti S.}
  \REVIEW{Phys. Rev. Lett.}{103}{2009}{118101}.
\bibitem{Boun}
  Note that this form of eigenvalue may be realized in the case of free boundary condition with $L\to\infty$ 
meaning that practically, $\rho_{n}=\infty$ for $n<1$ and $L<n$;
The case with large $L$, which gives Toeplitz matrix, could be still explicitly analyzed
in a perturbative manner with respect to $L^{-1}$ \cite{DGK}.
\bibitem{DGK}
   \Name{Dai H., Geary Z., \and Kadanoff L. P.}
  \REVIEW{J. Stat. Mech.}{P05012}{2009}.
\bibitem{eicir}
  \Name{Davis P. J.}
  \Book{Circulant Matrices 2nd ed.}
  \Publ{AMS Chelsea Publishing}
  \Year{1994}.
\bibitem{proof} Indeed, there is a rigorous statement that the set of initial conditions to converge
zigzag solutions is open and dense in $\mathbb{R}_{+}$ \cite{Smith}.
Although this statement might sound like contradicting the stable traveling waves found numerically,
this statement does not directly exclude such a possibility as implied also in \cite{Smith}.
The rigorous proof of the existence is beyond the scope of this paper.
\bibitem{Smith}
   \Name{Smith H. L.}
  \REVIEW{J. Math. Biol.}{25}{1987}{169}.
\bibitem{PSmith}
   \Name{Mallet-Paret J. \and Smith H. L.}
   \REVIEW{J. Dyn. Diff. Eqns.}{2}{1990}{367}.
\bibitem{Note1}
   Note that the case of feedback-loop network ${\mathbb F}^-$ with $h=2$
seems to show more bifurcations into quasiperiodic or, possibly,
chaotic solutions when $J$ is increased, as
observed in the case of the standard periodic boundary condition
to induce incommensurate spatial patterns \cite{JKP}.
\bibitem{Jensen3}
   \Name{Li W., Krishna S., Pigolotti S., Mitarai N., \and Jensen M. H.}
  \REVIEW{J. Theor. Biol.}{307}{2012}{205}.
\bibitem{Jensen2}
   \Name{Chakraborty S., Jensen M. H., Krishna S., Pers B. M., Pigolotti S., Sekara V., \and Semsey S.}
  \REVIEW{Phys. Rev. E}{86}{2012}{031905}.
\end{thebibliography}
\end{document}